\documentclass[12pt]{article}
 
\usepackage{scicite}   
\usepackage{times}     
\usepackage{setspace}  
\usepackage{lineno}    
 
\newenvironment{sciabstract}{%
  \begin{quote}\bfseries\boldmath\setlength{\emergencystretch}{2em}}{\end{quote}}
 
\usepackage{graphicx,amsfonts,amssymb,amsmath,hyperref}
\usepackage{lmodern}
\usepackage{booktabs}
 
\newif\ifhyper
\hypertrue
\ifhyper
\hypersetup{
   citecolor = {red},
   colorlinks = {true},
   urlcolor = {blue}
}
\fi
 
\newcommand{\beq}{\begin{equation}}
\newcommand{\eeq}{\end{equation}}
\newcommand{\beqa}{\begin{eqnarray}}
\newcommand{\eeqa}{\end{eqnarray}}

\newcommand{\avg}[1]{\langle #1 \rangle}
\newcommand{\lr}[1]{\left(#1\right)}
\usepackage{ragged2e}
\usepackage{etoolbox}
 
\begin{document}
 
\baselineskip 24pt   
 
\title{Pushing the Classical Frontier of 1D Fermi–Hubbard Quench Dynamics Beyond Current Quantum Simulations}
 
\author{
Roman Rausch,$^{1}$
Sukhbinder Singh,$^{2}$
Saeed S.\ Jahromi,$^{1,3}$\\[3pt]
Augustine Kshetrimayum,$^{1}$
Rom\'{a}n Or\'{u}s$^{1,3,4}$
\\[6pt]
\small $^{1}$Multiverse Computing, San Sebasti\'{a}n, Spain\\
\small $^{2}$Multiverse Computing, Toronto, Ontario, Canada\\
\small $^{3}$Donostia International Physics Center, San Sebasti\'{a}n, Spain\\
\small $^{4}$Ikerbasque Foundation for Science, Bilbao, Spain
}
 
\date{}
\maketitle
 
\begin{sciabstract}
Establishing quantum advantage requires comparison against the best achievable
classical simulation. The Q-CTRL team~\cite{qctrl2025fermihubbard} recently simulated quench dynamics of the
one-dimensional Fermi--Hubbard model on an IBM processor, completing a
$L=60$ evolution to time $t=6$ in under three minutes and claiming a $3000\times$ speedup
over classical Time-Dependent Variational Principle (TDVP) simulation at bond dimension $\chi=4096$. Their classical benchmark required
over 160 hours on a CPU cluster, failed to converge in the high-entanglement regime
$t\in[5.2,6]$, and left the most challenging window of the experiment unverified.
Here, we push the boundaries of classical simulation by exploiting the full
$\mathrm{U}(1)\times\mathrm{SU}(2)$ symmetry of the Fermi--Hubbard Hamiltonian
combined with GPU-accelerated tensor contractions. Reaching bond dimensions up to
$\chi\approx62{,}000$ on four NVIDIA H200 GPUs --- among the largest ever
achieved in TDVP simulations and fifteen times larger than Q-CTRL's classical
baseline --- we achieve fully converged results across the entire simulation window,
including rigorous certification of the previously unresolved high-entanglement regime
$t\in[5.2,6]$. We further advance the classical frontier to $t=7$, which
lies beyond the quantum hardware experiment and any previously verified
classical evolution of the full wavefunction. At the bond dimension comparable to Q-CTRL's best classical run,
our GPU implementation completes in $\sim\!100$ minutes, directly reducing the claimed
$3000\times$ quantum advantage to $\sim\!36\times$. These results substantially narrow the quantum–classical performance gap and establish a new standard for tensor-network benchmarking of large-scale quantum simulations.
\end{sciabstract}
 
 
\section{Introduction}
\label{sec:intro}
 
Quantum processors have grown rapidly in scale and quality, and the field is now
shifting from demonstrating that deep circuits can be run to asking whether they
deliver genuine scientific value~\cite{nisqPreskill, daley2022practical}. Quench
dynamics of the one-dimensional Fermi--Hubbard model (FHM) has emerged as a
leading testbed for this question~\cite{ChowdhuryFHMQutility2026, qctrl2025fermihubbard}:
the model supports spin-charge separation and Luttinger liquid physics, its
entanglement grows rapidly under time evolution, and it maps naturally onto
superconducting qubits via the Jordan--Wigner
transformation~\cite{cade2020strategies, stanisic2022observing}. Rigorous classical
verification remains essential, however. Gate errors accumulate over long circuits,
and without a reliable classical reference it is impossible to determine whether a
quantum device faithfully tracks the true dynamics. Matrix Product State (MPS)
methods, and in particular the Time-Dependent Variational Principle~\cite{tdvp_verstraetePRL2011, tdvp_verstraete2016, paeckel2019time}, are
the leading classical tool for this purpose in one dimension, with accuracy
controlled by the bond dimension $\chi$. Notably, in several recent quantum
advantage experiments tensor-network simulations have matched or surpassed the
quantum hardware output~\cite{TindallBPPRX2024, PatragPEPS2024, XiangPEPOIBM2023,
TindallScience2026}, underscoring the need to deploy the most capable classical
methods before any advantage claim is accepted~\cite{KshetrimayumTNperspective2026}.

The largest FHM quench simulation reported to date was performed by the Q-CTRL
team~\cite{qctrl2025fermihubbard}, who evolved a half-filled N\'{e}el state on
$L=60$ fermionic sites (120 qubits) to time $t=6$ using an IBM Heron
superconducting processor, completing in 2 minutes 46 seconds of QPU time. For
the classical benchmark they used $\mathrm{U}(1)\times\mathrm{U}(1)$-symmetric
TDVP at bond dimensions up to $\chi=4096$ via the ITensor
library~\cite{fishman2022itensor}, requiring over 160 hours on a CPU cluster.
This benchmark agreed with the quantum hardware for $t\lesssim 5.2$, but diverged
beyond that point, leaving the high-entanglement window $t\in[5.2,6]$ unverified.
Comparing the CPU runtime to the bare QPU time, the authors claimed a $3000\times$
quantum speedup.

Here we push the boundaries of classical simulation to directly address this
verification gap. By exploiting the full $\mathrm{U}(1)\times\mathrm{SU}(2)$
symmetry of the half-filled FHM --- going beyond the $\mathrm{U}(1)\times\mathrm{U}(1)$
subgroup used by Q-CTRL --- combined with GPU-accelerated tensor contractions, we
reach bond dimensions up to $\chi\approx62{,}000$ on four NVIDIA H200
GPUs, among the largest ever achieved in TDVP simulations. At the bond dimension
comparable to Q-CTRL's best classical run, our implementation completes in
$\sim\!100$ minutes, directly reducing the claimed $3000\times$ quantum speedup
to $\sim\!36\times$. Scaling to larger bond dimensions, we achieve fully converged
results across the entire Q-CTRL window --- providing the first rigorous classical
certification of the high-entanglement regime $t\in[5.2,6]$ --- and extend the
simulation to $t=7$, beyond both the reach of the quantum hardware and any
previously verified classical benchmark.
 
\section{Model and Simulation Setup}
\label{sec:model}

The 1D Fermi--Hubbard model is defined by the Hamiltonian
\beq
H = -t_h \sum_{\langle i,j\rangle,\sigma}\!\left(c^\dagger_{i\sigma}c_{j\sigma}+\mathrm{h.c.}\right) + U\sum_i n_{i\uparrow}n_{i\downarrow},
\label{eq:FHM}
\eeq
where $c^\dagger_{i\sigma}$ ($c_{i\sigma}$) creates (annihilates) a fermion of spin
$\sigma\in\{\uparrow,\downarrow\}$ on site $i$, and $n_{i\sigma}=c^\dagger_{i\sigma}c_{i\sigma}$
is the number operator. Following Ref.~\cite{qctrl2025fermihubbard}, we set $t_h=1$ as
the unit of energy and consider a chain of $L=60$ sites with open boundary conditions,
matching the largest system studied in their quantum simulation.

At half-filling, the spectral properties of the FHM depend on the sign of $U$. For
repulsive interactions ($U>0$), a finite charge gap opens (the Mott gap), while the
spin sector remains gapless, giving rise to spin-charge separation and
Tomonaga--Luttinger liquid behavior. For attractive interactions ($U<0$), we are in
the Luther--Emery liquid regime where the roles are exchanged: a spin gap develops due
to singlet pair formation, while the charge sector remains gapless. The Q-CTRL
experiment~\cite{qctrl2025fermihubbard} operates in this latter regime with $U=-2$,
using 30 first-order Trotter steps of size $\Delta t=0.2$. The quench dynamics reveal
holons and spinons propagating at distinct velocities --- a signature of spin-charge
separation.

\subsection{Symmetries: the key to acceleration}

The FHM conserves total fermion number $N$ and $z$-component of spin $S^z$, giving a
$\mathrm{U}(1)_\mathrm{charge}\times\mathrm{U}(1)_\mathrm{spin}$ symmetry --- the
combination exploited by Q-CTRL~\cite{qctrl2025fermihubbard}. The full symmetry group, however, is larger: $\mathrm{SO}(4)\cong\mathrm{SU}(2)_\mathrm{spin}\times\mathrm{SU}(2)_\mathrm{charge}$,
where the charge $\mathrm{SU}(2)$~\cite{Anderson1958RPA, Zhang1990Pseudospin}
interconverts empty and doubly occupied sites. In the Q-CTRL experiment, the quantum simulation begins from the
half-filled N\'{e}el state
\begin{equation}\label{eq:neel}
|\!\uparrow\downarrow\uparrow\downarrow\cdots\rangle
=
\left(\prod_{i\in A} c_{i\uparrow}^{\dagger}\right)
\left(\prod_{j\in B} c_{j\downarrow}^{\dagger}\right)
|0\rangle,
\end{equation}
where $A$ and $B$ denote the two sublattices. This state has total pseudospin $T_\mathrm{tot}=0$ and $S^z=0$, but not a well-defined total spin $S_\mathrm{tot}=0$, thus breaking the full symmetry to a
$\mathrm{U}(1)_\mathrm{spin}\times\mathrm{SU}(2)_\mathrm{charge}$ symmetry throughout the dynamics.
Exploiting this symmetry (larger than $\mathrm{U}(1)_\mathrm{spin}\times\mathrm{U}(1)_\mathrm{charge}$ considered in \cite{qctrl2025fermihubbard}), which groups tensor indices into multiplets via the
Wigner--Eckart theorem~\cite{symtn1, symtn2, symtn3, symtn4}, reduces the effective
number of independent tensor components by a factor of
$\mathcal{O}(\chi^{1/3})$--$\mathcal{O}(\chi^{1/2})$~\cite{McCulloch_SU2_2007,symtn3,symtn4, weichselbaum}.
Since TDVP cost scales as $\chi^3$, this alone yields a large speedup at fixed
accuracy. Technical details of the symmetry decomposition and bond-dimension
taxonomy are given in the Supplementary Materials.

\section{GPU-accelerated \texorpdfstring{$\mathrm{U}(1)\times\mathrm{SU}(2)$}{U(1)xSU(2)}-symmetric TDVP}
\label{sec:methods}

The classical TDVP baseline in Ref.~\cite{qctrl2025fermihubbard} was run using the
ITensor Julia library~\cite{fishman2022itensor} on an AWS \texttt{c7i.8xlarge}
instance (32\,vCPU, 64\,GB RAM), employing $\mathrm{U}(1)\times\mathrm{U}(1)$
symmetry. At bond dimension $\chi=4096$, this required over 160 hours to reach $t=6$.
The claimed $3000\times$ quantum advantage was obtained by comparing this runtime to
the $\approx\!166$\,seconds of bare QPU execution time, excluding readout error
mitigation, decay recovery, circuit compilation, and post-processing. Our
implementation improves upon this baseline on three fronts; full algorithmic details
are provided in the Supplementary Materials.

\textit{Symmetry.} Encoding the $\mathrm{U}(1)_\mathrm{spin}\times\mathrm{SU}(2)_\mathrm{charge}$
symmetry --- absent in the ITensor baseline --- into the MPS tensors via the
Wigner--Eckart theorem~\cite{McCulloch_SU2_2007, symtn2, symtn3, symtn4, weichselbaum} groups all
$2T+1$ components of each pseudospin-$T$ multiplet into a single reduced matrix
element. In practice we find $\chi\approx 2.4\,\chi_{\mathrm{SU(2)}}$,
so the computational cost scales with the smaller $\chi_{\mathrm{SU(2)}}$ while the
expressive power matches $\chi$.

\textit{GPU acceleration.} We implemented the algorithm in
PyTorch~\cite{symtngpu}, exploiting GPU GEMM throughput for the dominant
environment-update and Krylov-exponentiation steps. To preserve arithmetic
intensity from the block-diagonal symmetry structure --- a known challenge for
GPU-accelerated symmetric tensor networks~\cite{iTensorSym} --- tensors of equal
block size are dispatched as batched-GEMM calls. At large bond dimensions, memory
becomes the binding constraint; we use \texttt{float32} arithmetic, validated
against \texttt{float64} runs (see SM). For our largest simulations
($\chi\approx62{,}000$) we distribute batched matrix multiplications across
four H200 GPUs using a greedy load-balancing scheme, with SVDs performed on the CPU.

\textit{Adaptive 1-site/2-site switching.} We begin with 2-site TDVP updates to
allow the bond dimension to grow freely, then switch to 1-site updates once it
saturates, eliminating per-step SVD overhead entirely. This yields a
$4$--$10\times$ wall-clock speedup for the later time steps relative to 2-site
updates throughout, as used by Q-CTRL. The SVD truncation threshold
$\epsilon_\mathrm{tol}=10^{-4}$ --- looser than the $10^{-8}$ used by Q-CTRL ---
delays bond-dimension saturation and reduces average cost per step while delivering
converged observables; an error analysis is given in the Supplementary Materials.

\section{Results}
\label{sec:results}

\subsection{Convergence and verification of \texorpdfstring{$t\in[5.2,6]$}{t in [5.2,6]} regime}

\begin{figure*}[htbp]
\centering
\includegraphics[width=\textwidth]{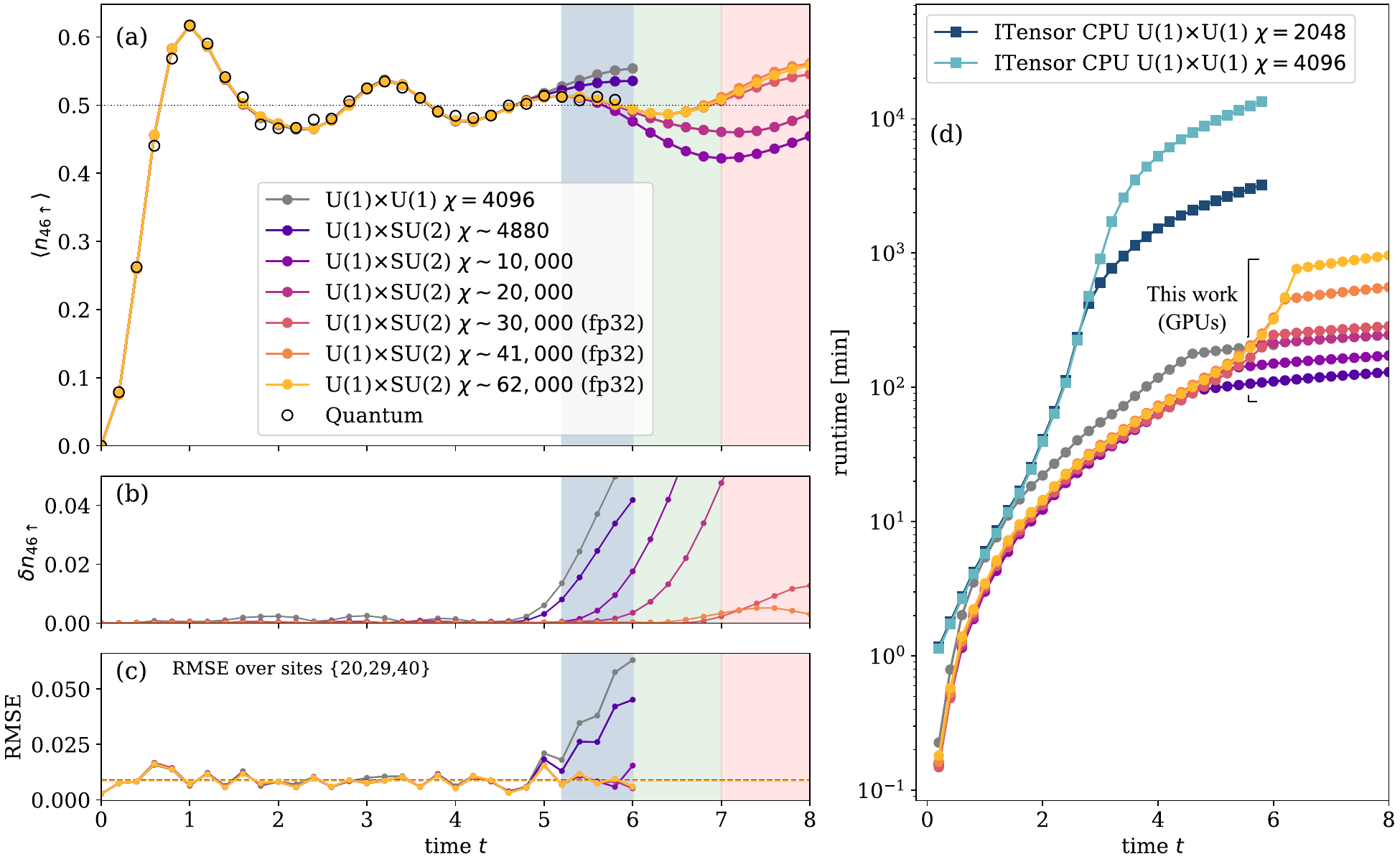}
\caption{Main results.
(a) TDVP time evolution of $\avg{n_{46\uparrow}}$ up to $t=8$ for $L=60$ fermions
with $\epsilon_{\text{tol}}=10^{-4}$. Our U(1)$\times$U(1) curve at $\chi=4096$
reproduces the Q-CTRL iTensor calculation and confirms non-convergence beyond
$t\approx5$. Our U(1)$\times$SU(2) TDVP with $\chi\approx30{,}000$ achieves
convergence in $t\in[5.2,6]$ (shaded blue) and our results are reliable in
$t\in[6,7]$ (shaded green), extending beyond the quantum simulation window.
(b) Absolute deviation of $\avg{n_{46\uparrow}}$ from the $\chi\approx62{,}000$
reference at each bond dimension.
(c) RMSE of our TDVP data against Q-CTRL's published values of
$\avg{n_{20\uparrow}}$, $\avg{n_{29\uparrow}}$, and $\avg{n_{40\uparrow}}$.
(d) Wall-clock runtimes: ITensor TDVP on CPU cluster~\cite{qctrl2025fermihubbard}
vs.\ our GPU implementation (same legend as (a)).
}
\label{fig:main}
\end{figure*}

Figure~\ref{fig:main}(a) shows the TDVP time evolution of $\langle n_{46\uparrow}\rangle$
for $L=60$ fermions up to $t=8$ (converged until $t=7$, see Sec.~\ref{sec:convt7}) at truncation threshold $\epsilon_\mathrm{tol}=10^{-4}$.
Our $\mathrm{U}(1)\times\mathrm{U}(1)$ result at $\chi=4096$ reproduces the Q-CTRL
iTensor calculation, confirming that TDVP at this bond dimension fails to converge
beyond $t\approx5$. By contrast, our $\mathrm{U}(1)\times\mathrm{SU}(2)$ simulation
at $\chi\approx30{,}000$ achieves convergence throughout the high-entanglement
window $t\in[5.2,6]$ (shaded blue). Figure~\ref{fig:main}(b) shows the absolute
deviation from the $\chi\approx62{,}000$ reference; it falls below $0.01$ at
$\chi\approx20{,}000$ for all $t\leq6$. Figure~\ref{fig:main}(c) shows the
RMSE (see SM for definition) between our TDVP expectation values and the Q-CTRL quantum hardware data for
three representative sites. The RMSE diverges at $\chi\approx4{,}880$ but
remains small for $\chi>10{,}000$ throughout $t\in[0,6]$, providing
rigorous classical certification of the entire Q-CTRL window including the previously
unresolved high-entanglement regime.

\subsection{Wall-time performance and revised quantum advantage}

Figure~\ref{fig:main}(d) and Table~\ref{tab:runtime_summary} compare runtimes. At
$\chi=4096$, the Q-CTRL ITensor simulation on 32\,vCPUs requires over 160\,h; our
GPU-accelerated $\mathrm{U}(1)\times\mathrm{U}(1)$ implementation at the same bond
dimension completes in $\approx200$\,min. Switching to $\mathrm{U}(1)\times\mathrm{SU}(2)$
symmetry at comparable expressive power ($\chi\approx4{,}880$) reduces this to
$\approx100$\,min --- a $\sim\!100\times$ speedup over the ITensor CPU baseline and,
when compared against the QPU time of 166\,s, a revised quantum advantage of
$\sim\!36\times$ rather than the claimed $3000\times$. Full runtimes across all bond
dimensions and tolerances are given in Table~\ref{tab:runtime_to_t7} of the Supplementary Materials.

\begin{table}[t]
  \centering
  \caption{Wall-clock runtime to $t=6$ at $\epsilon_\mathrm{tol}=10^{-4}$. The Q-CTRL
  QPU entry gives bare processor time. Full runtimes across all tolerances are in
  Table~\ref{tab:runtime_to_t7}.}
  \begin{tabular}{l l r r}
    \toprule
    Implementation & $\chi$ & $N_\mathrm{GPU}$ & Runtime \\
    \midrule
    Q-CTRL QPU~\cite{qctrl2025fermihubbard} & --- & --- & 2\,m\,46\,s \\
    Q-CTRL ITensor U(1)$\times$U(1)~\cite{qctrl2025fermihubbard} & 4\,096 & 32\,vCPU & $>$160\,h \\
    \midrule
    This work with U(1)$\times$U(1), fp64 & 4\,096 & 1 & 3\,h\,27\,m \\
    This work with U(1)$\times$SU(2), fp32 & 4\,880 & 1 & 1\,h\,40\,m \\
    This work with U(1)$\times$SU(2), fp32 & 30\,000 & 1 & 4\,h\,06\,m \\
    This work with U(1)$\times$SU(2), fp32 & 62\,000 & 4 & 5\,h\,21\,m \\
    \bottomrule
  \end{tabular}
  \label{tab:runtime_summary}
\end{table}

\subsection{Full spatiotemporal benchmark}

\begin{figure*}[htbp]
\centering
\includegraphics[width=0.9\textwidth]{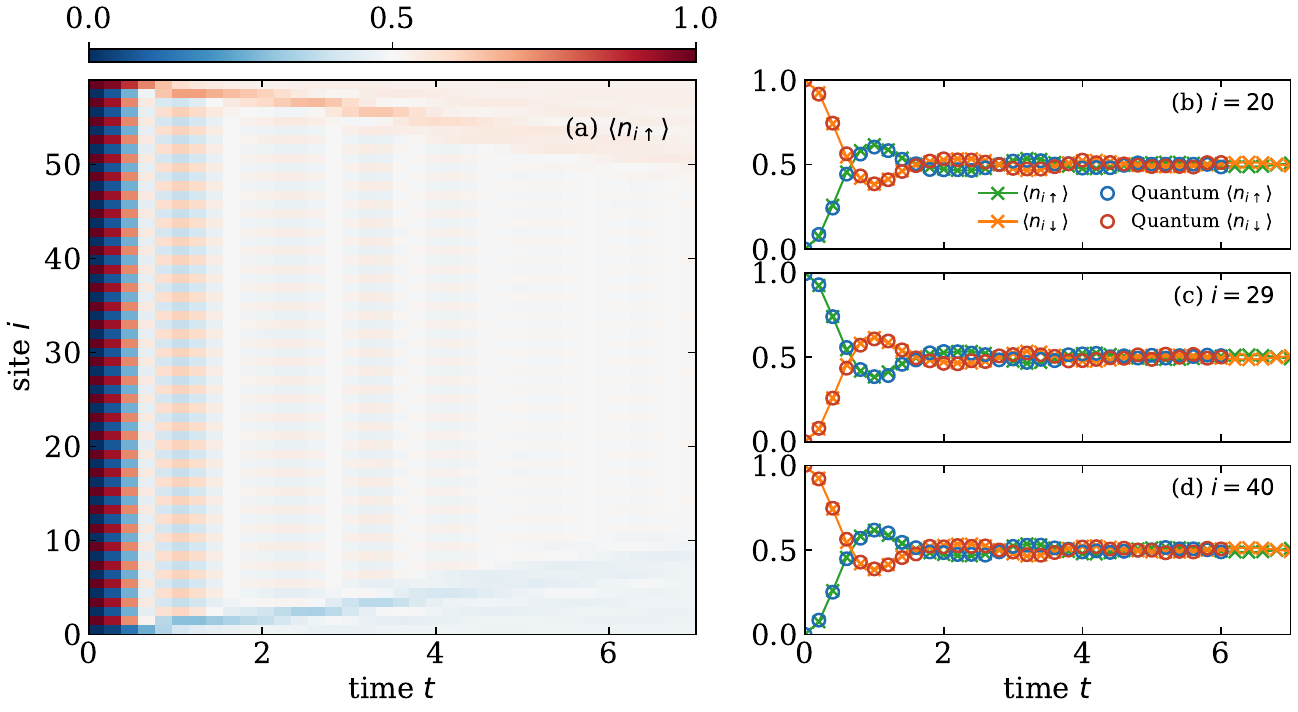}
\caption{Classical benchmarks for the Q-CTRL quantum simulation results of Fig.~3 in
Ref.~\cite{qctrl2025fermihubbard}.
(a) Expectation value $\langle n_{i\uparrow}\rangle$ (up-spin density at site $i$)
as a function of evolution time $t$, computed with our
$\mathrm{U}(1)\times\mathrm{SU}(2)$ TDVP at $\chi\approx62{,}000$
(\texttt{float32}, $\epsilon_{\text{tol}}=10^{-4}$). Results already converged for $t \leq 6$ using $\chi = 30,000$ (Fig~\ref{fig:res_heatmap_12288}).
(b)--(d) Site-resolved values at $i=20$, $29$, $40$ compared directly with the
Q-CTRL quantum simulation results.}
\label{fig:heatmap}
\end{figure*}

Figure~\ref{fig:heatmap} shows the site-resolved up-spin density $\langle n_{i\uparrow}\rangle$
computed with our $\mathrm{U}(1)\times\mathrm{SU}(2)$ TDVP at $\chi\approx62{,}000$
(\texttt{float32}). This constitutes a converged classical benchmark for the full
spatiotemporal density profile of Ref.~\cite{qctrl2025fermihubbard} (Fig.~3 therein),
and extends it to $t=7$. For comparison, the $\mathrm{U}(1)\times\mathrm{U}(1)$
result at $\chi=4096$ (Fig.~\ref{fig:profile_u1xu1_4096}) diverges visibly at late times, confirming that
a larger bond dimension is essential. Results at $t\leq6$ are already converged
at $\chi\approx30{,}000$ (\texttt{float32}), see Fig.~\ref{fig:res_heatmap_12288}.

\subsection{Extension to \texorpdfstring{$t=7$}{t=7}, beyond the quantum hardware} \label{sec:convt7}

Figure~\ref{fig:main}(a) also shows results for $t\in[6,7]$ (shaded green), a window
beyond both the Q-CTRL quantum simulation and any previously verified classical
benchmark. The TDVP converges in the range $\chi\approx40{,}000$--$62{,}000$;
a shallow minimum around $t\approx6.4$ is cleanly resolved. Beyond $t=7$, bond-dimension
spread grows and we do not certify the dynamics in that regime.

\section{Conclusions}
\label{sec:conclusions}

We have advanced the classical benchmark for the Q-CTRL Fermi--Hubbard quench
simulation~\cite{qctrl2025fermihubbard} on four fronts. First, we certify the quantum
simulation across the full window $t\in[0,6]$, including the high-entanglement regime
$t\in[5.2,6]$ that the Q-CTRL classical benchmarks could not resolve. Second, our
$\mathrm{U}(1)\times\mathrm{SU}(2)$ GPU implementation at comparable bond dimension
($\chi\approx4{,}880$) completes in $\approx\!100$\,minutes --- a $\sim\!100\times$
speedup over the ITensor CPU baseline --- which directly revises the claimed
$3000\times$ quantum advantage down to $\sim\!36\times$ when compared against the
bare QPU time, and calls for a reassessment of that figure. Third, at
$\chi\approx62{,}000$ --- among the largest TDVP bond dimensions achieved to
date --- we provide the first converged spatiotemporal reference for the full $L=60$
density heatmap. Fourth, we
extend the classical frontier to $t=7$, beyond the reach of the quantum hardware itself.
Taken together, these results show that combining symmetry exploitation, GPU acceleration,
and adaptive truncation substantially pushes the classical simulability frontier.

The timings reported here are for full wavefunction evolution. Observable-specific
methods --- such as Majorana propagation~\cite{miller2025majorana} or Heisenberg-picture
operator evolution --- can in principle extend the simulated timescale further for
particular observables of interest, and our symmetry and hardware optimizations apply
equally to operator evolution.

Further progress is likely on both hardware and algorithmic fronts. Larger GPU clusters
and more efficient GPU-adapted algorithms~\cite{legeza1,legeza2,legeza3,nvidiaTN} will reduce runtimes at fixed bond dimension, while algorithms based on variational
compression~\cite{schollwock_MPSreview_2011, vandamme2021tangent} or global subspace
expansion~\cite{yang2020time, dunnett2021matrix} may achieve more faithful approximations
at the same $\chi$. The classical frontier for Fermi--Hubbard dynamics has moved ---
the question is how much further it can go.

\section*{Acknowledgments}
 
We thank Matthias Peschke for discussions as well as Multiverse Computing for support and computational resources.  

\paragraph*{Author contributions:}
R.O.\ and S.S.J.\ conceived and directed the project. R.R.\ and S.S.\ developed the
algorithm; R.R.\ implemented it in PyTorch and performed the numerical experiments.
R.R., S.S., S.S.J., and A.K.\ designed the experiments and analysed the data. All
authors contributed to writing the manuscript.

\paragraph*{Competing interests:}
The authors declare no competing interests.

\paragraph*{Data, code and materials availability:}
Data, code and materials availability: The data and other materials used in this study are available from the corresponding author upon reasonable request, subject to institutional and commercial constraints.

\clearpage 

%
\bibliographystyle{sciencemag}
\bibliography{references}


\clearpage

\setcounter{figure}{0}
\setcounter{table}{0}
\setcounter{section}{0}
\renewcommand{\thefigure}{S\arabic{figure}}
\renewcommand{\thetable}{S\arabic{table}}
\renewcommand{\thesection}{S\arabic{section}}
\renewcommand{\thesubsection}{S\arabic{section}.\arabic{subsection}}

\begin{center}
{\Large\textbf{Supplementary Materials}}\\[6pt]
{\large\textbf{Pushing the Classical Frontier of 1D Fermi–Hubbard Quench Dynamics Beyond Current Quantum Simulations}}\\[4pt]
{\normalsize
Roman Rausch,$^{1}$
Sukhbinder Singh,$^{2}$
Saeed S.\ Jahromi,$^{1,3}$
Augustine Kshetrimayum,$^{1}$
Rom\'{a}n Or\'{u}s$^{1,3,4}$
}\\[4pt]
{\small
$^{1}$Multiverse Computing, San Sebasti\'{a}n, Spain\\
$^{2}$Multiverse Computing, Toronto, Ontario, Canada\\
$^{3}$Donostia International Physics Center, San Sebasti\'{a}n, Spain\\
$^{4}$Ikerbasque Foundation for Science, Bilbao, Spain}
\end{center}
\vspace{6pt}

\section{The Q-CTRL classical benchmark details}

The Q-CTRL team's primary experiment reports the quench dynamics of the 1D FHM with attractive coupling $U = -2$, evolving from two initial states: a
half-filled N\'{e}el state for $L = 60$ sites, Eq.~\ref{eq:neel}, and a
N\'{e}el state with a single central vacancy for $L = 31$ sites.
The time evolution reaches $t = 6$ via 30 first-order Trotter steps of size
$\Delta t = 0.2$, implemented as a circuit of $9{,}057$ two-qubit gates at depth 152.
The site-resolved spin-up occupations $\langle n_{i\uparrow}\rangle$ measured by the
quantum hardware reveal clear signatures of spin--charge separation in the vacancy
experiment: spinon and holon excitations propagate at distinct velocities, consistent
with Luttinger liquid theory.
 
As a classical benchmark, TDVP simulations with $\mathrm{U}(1)\times\mathrm{U}(1)$
symmetry were performed at bond dimensions
$\chi\in\{64,128,256,512,1024,4096\}$ using the ITensor library~\cite{fishman2022itensor}.
Quantum--classical agreement was quantified by the root-mean-square error (RMSE) over all
$2L = 120$ spin orbitals,
\begin{equation}
    \mathrm{RMSE} = \left(\frac{1}{2L}\sum_{i,\sigma}
    \bigl(\langle n_{i\sigma}\rangle_\mathrm{QC}
    - \langle n_{i\sigma}\rangle_\mathrm{TDVP}\bigr)^2\right)^{1/2}.
\end{equation}
The RMSE remains small for $t \lesssim 5.2$ but grows rapidly at later times
even at $\chi = 4096$, indicating that this bond dimension is insufficient to resolve
the dynamics in the high-entanglement regime.
Pauli path propagation (PPP) was also evaluated as a second classical baseline and
similarly failed to track the quantum evolution at late times.
On the computational side, TDVP at $\chi = 4096$ required over 160 hours of wall-clock
time to reach $t = 6$, while the quantum hardware completed the full simulation
--- including readout error mitigation and decay recovery --- in $\approx 4$min 25s;
this figure excludes circuit compilation, data retrieval from IBM Cloud, and post-processing.

\section{Materials and Methods}

\subsection{Exploiting the full \texorpdfstring{$\mathrm{U}(1) \times \mathrm{SU}(2)$}{U(1) x SU(2)} symmetry}

The FHM possesses exact symmetries whose exploitation is central to our simulation
strategy. For any $U$, the total fermion number $N = N_{\uparrow} + N_{\downarrow}$
and the $z$-component of total spin $S^z = \tfrac{1}{2}(N_{\uparrow} - N_{\downarrow})$
are separately conserved, giving a $\mathrm{U}(1)_{\mathrm{charge}} \times
\mathrm{U}(1)_{\mathrm{spin}}$ symmetry --- the combination used in the Q-CTRL
classical benchmark~\cite{qctrl2025fermihubbard}. However, each $\mathrm{U}(1)$
is embedded in a larger $\mathrm{SU}(2)$, and the full symmetry group is
$\mathrm{SO}(4) \cong \mathrm{SU}(2)_{\mathrm{spin}} \times \mathrm{SU}(2)_{\mathrm{charge}}$.
The spin $\mathrm{SU}(2)$ is generated by the total spin operators $S^\pm$ and $S^z$.
The charge $\mathrm{SU}(2)$~\cite{Anderson1958RPA, Zhang1990Pseudospin} --- not
exploited in Ref.~\cite{qctrl2025fermihubbard} --- is generated by the pseudospin
operators
\begin{equation}
    T^+ = \sum_{i=1}^{L} (-1)^i \, c^\dagger_{i\uparrow} c^\dagger_{i\downarrow},
    \;
    T^- = \left(T^+\right)^\dagger,
    \;
    T^z = \frac{N - L}{2},
\end{equation}
which interconvert empty and doubly occupied sites and satisfy the $\mathrm{SU}(2)$
algebra
\begin{equation}
    \left[T^+,\, T^-\right] = 2T^z, \qquad \left[T^z,\, T^\pm\right] = \pm\,T^\pm.
\end{equation}
These operators commute with $H$,
\begin{equation}
    \left[H,\, T^\pm\right] = 0, \qquad \left[H,\, T^z\right] = 0,
\end{equation}
and are therefore conserved throughout the time evolution. The alternating phase
$(-1)^i$ in $T^\pm$ requires a bipartite lattice, which is satisfied by the open
chain studied here.

The quantum simulation begins from the half-filled N\'{e}el state, which lies in the
$S^z = 0$, $T_{\mathrm{tot}} = 0$\footnote{$T_{\mathrm{tot}}$ is the total
pseudospin quantum number, defined via the Casimir
$(T^x)^2 + (T^y)^2 + (T^z)^2 = T_{\mathrm{tot}}(T_{\mathrm{tot}}+1)$,
where $T^\pm = T^x \pm iT^y$. At half-filling ($N=L$), $T^z=0$.} sector, but does not have a well-defined total spin $S_{\mathrm{tot}} = 0$.
Both quantum numbers $S^z = 0$, $T_{\mathrm{tot}} = 0$ are conserved under time
evolution by $H$. Exploiting the resulting
$\mathrm{U}(1)_{\mathrm{spin}}\times\mathrm{SU}(2)_{\mathrm{charge}}$ symmetry
is the key ingredient that accelerates our TDVP simulation beyond the Q-CTRL benchmark.\footnote{At half-filling the model also possesses the
particle-hole symmetry $c_{i\sigma} \to (-1)^i c^\dagger_{i\sigma}$, equivalent to
$T^z_i \to -T^z_i$, which yields no further compression when the full charge
$\mathrm{SU}(2)$ is already enforced. The open boundary conditions (OBC) reduce
translational symmetry to spatial inversion; while this could in principle be
exploited, it offers no practical advantage for MPS-based codes, which are already
highly efficient under OBC.}

Under the action of the $\mathrm{U}(1)\times\mathrm{SU}(2)$ symmetry each MPS bond space
$\mathbb{V}^{[i]}$ (Eq.~\ref{eq:MPS}) decomposes as
\begin{equation}\label{eq:symdecomp}
    \mathbb{V}^{[i]} = \bigoplus_{n,j}
    \mathbb{D}^{[i]}_{n,j} \otimes \mathbb{V}^{[i]}_{n,j},
\end{equation}
where $\mathbb{V}_{n,j}$ is the $(2j+1)$-dimensional irrep labeled by
particle number $n$ and spin $j$, and $\mathbb{D}_{n,j}$ is the degeneracy
space of dimension $d_{n,j}$.%
\footnote{The symmetry acts as the identity on $\mathbb{D}_{n,j}$; all
variational degrees of freedom reside there.} As a concrete example, Table~\ref{tab:irrep_degeneracies} lists the sector
degeneracies at $t=6$ for our $\chi_{\mathrm{SU(2)}}=12{,}288$
($\chi\approx 30{,}000$) simulation.

The decomposition Eq.~\ref{eq:symdecomp} gives rise to three bond-dimension parameters:
\begin{align}
\chi &= \sum_{n,j} d_{n,j}(2j+1), \\
d^{\mathrm{max}}_{\mathrm{SU(2)}} &= \max_{n,j}\, d_{n,j}, \\
\chi_{\mathrm{SU(2)}} &= \sum_{n,j} d_{n,j}.
\end{align}
The \emph{effective} bond dimension $\chi$ is the total dimension of the bond space,
directly comparable to the bond dimension of the same MPS in a non-symmetric basis.
The \emph{total symmetric} bond dimension $\chi_{\mathrm{SU(2)}}$ is the truncation
parameter we impose in our implementation. The \emph{maximum} degeneracy
$d^{\mathrm{max}}_{\mathrm{SU(2)}}$ most directly governs computational cost: working
in the irrep basis reduces the SVD cost from $\mathcal{O}(\chi^3)$ to
$\mathcal{O}((d^{\mathrm{max}}_{\mathrm{SU(2)}})^3)$. In practice we find
$\chi\approx 2.4\,\chi_{\mathrm{SU(2)}}$. For $\mathrm{U}(1)\times\mathrm{U}(1)$
symmetry, $\chi = \chi_{\mathrm{U(1)}}$. Throughout this work, all quoted bond
dimensions refer to $\chi$.

\subsection{Adaptive TDVP algorithm}

The time-dependent variational principle~\cite{tdvp_verstraetePRL2011,tdvp_verstraete2016} (TDVP) is the leading method for simulating the time evolution of 1D quantum systems beyond the reach of exact diagonalization. On a lattice with open boundary conditions, an MPS with bond dimension $\chi$ approximates the many-body wavefunction as
\beq
|\psi\rangle = \sum_{\{s_i\}} A^{s_1}_1 A^{s_2}_2 \cdots A^{s_L}_L |s_1 s_2 \cdots s_L\rangle,
\label{eq:MPS}
\eeq
where $s_i \in \{1,\ldots,d\}$ labels a basis state of the local $d$-dimensional Hilbert space, $A^{s_i}_i: \mathbb{V}^{[i-1]} \rightarrow  \mathbb{V}^{[i]}$ are matrices acting on bond spaces $\{\mathbb{V}^{[i]}\}$ with $\dim(\mathbb{V}^{[i]}) = \chi_i$. The maximum value $\max_i\{\chi_i\}$ is the bond dimension of the MPS.
 
TDVP projects the Schr\"odinger equation onto the MPS tangent space and integrates the
resulting equations of motion via DMRG-like sweeps over the lattice. At each sweep step, one or two
MPS tensors are locally updated by exponentiating the effective Hamiltonian of the
corresponding block via a Krylov-based matrix exponential; the Krylov subspace dimension
is chosen adaptively~\cite{Lubich2008QuantumClassicalMD}, reducing unnecessary iterations
at short effective time steps. The dominant cost per time step
scales as $\mathcal{O}(L \chi^3 d)$, where $d$ is the local Hilbert space dimension.
 
The key distinction between the 1-site and 2-site TDVP variants lies in how the bond dimension is controlled.
In the 2-site algorithm, two adjacent tensors are merged and the bond between them is updated via SVD,
allowing the bond dimension to grow adaptively during the sweep.
The 1-site algorithm, by contrast, keeps the bond dimension fixed, requiring it to be set
large enough at initialisation to accommodate the entanglement at the end of the simulation.
For quench dynamics --- where entanglement grows rapidly --- the 2-site algorithm is the more
practical choice, since it avoids over-provisioning bond dimension early in the evolution. When symmetries are exploited (see next section), the 2-site algorithm also naturally provides a suitable distribution of the total bond dimension over the sparse blocks of the MPS tensors.
 
The Q-CTRL team employed 2-site TDVP throughout their simulation. We instead adopt an
adaptive hybrid scheme: we begin with 2-site updates to allow the bond dimension to grow
freely, then switch to 1-site updates once it has saturated. This yields a substantial
runtime speedup --- empirically factors of $4$--$10$ for the subsequent time steps. The gain
arises because the 2-site algorithm must perform an SVD of the merged two-site tensor
(of size $d^2\chi^2$; $d = 4$ for the FHM) at every sweep step, an operation that becomes
pure overhead once bond growth has ceased. Switching to 1-site updates, which propagate
tensors of size $d\chi^2$, eliminates this cost entirely for the remainder of the simulation.
 
Since our implementation is independent of ITensor, we can apply this hybrid scheme natively
without any manual partitioning of the time evolution. The Q-CTRL team did not report any
such adaptation, and we assume their simulation used the 2-site algorithm throughout.

\subsection{GPU Acceleration}
We implemented our $\mathrm{U}(1)\times\mathrm{SU}(2)$-symmetric TDVP algorithm in PyTorch with targeted low-level
optimizations to leverage GPU acceleration~\cite{symtngpu}. Similar approaches have been recently reported in quantum chemistry~\cite{legeza1,legeza2,legeza3,nvidiaTN}.
 
The dominant computational bottleneck in TDVP is the contraction of MPS tensors with the
matrix product operator (MPO) representation of the Hamiltonian $H$. More specifically, the dominant computational cost arises due to the updates of the left and right environments and the local Krylov exponentiation at each bond. These operations reduce to dense matrix multiplications involving tensors whose leading dimension scales as $\chi^2 d W$, where $W$ is the MPO bond dimension ($W = 4$ for the FHM with $\mathrm{U}(1)\times\mathrm{SU}(2)$). At the large
matrix sizes reached in our simulations, GPU GEMM\footnote{The routine for GEneral Matrix Multiplications in the cuBLAS library.} throughput substantially exceeds that of
CPUs, making GPU execution a natural fit.

GPU-accelerating \emph{symmetric} tensor network algorithms presents an additional challenge.
The computational gain from symmetry arises by decomposing large tensors into block-diagonal
form, replacing a single large matrix multiplication with a collection of smaller ones. On a
GPU, however, this block structure can severely dilute arithmetic intensity: when individual
blocks are small, the cost of data movement dominates over floating-point work, negating the
GPU advantage---a limitation also noted in the ITensor library documentation~\cite{iTensorSym}.
We mitigate this by \emph{batching}: blocks of equal size are grouped and dispatched
together as a single batched-GEMM call, replacing many small independent kernel launches
with a shorter sequence of higher-occupancy operations. At large bond dimensions, the runtime is dominated by the contractions of the largest batches, which are substantially accelerated by GPUs.

At sufficiently large bond dimension, memory becomes the binding constraint on the GPU.
We address this by reducing floating-point precision from \texttt{float64} to \texttt{float32},
which also yields a modest wall-clock speedup of $6$--$23\%$ depending on bond dimension.
 
Floating-point precision in classical simulations must be commensurate with the accuracy of the hardware being benchmarked. The Q-CTRL quantum circuits are wide and deep, and the resulting noise floor of the QPU substantially exceeds the rounding errors introduced by single-precision (float32) arithmetic. Using double precision (float64) would therefore provide no meaningful improvement in benchmark fidelity while significantly increasing memory and compute costs. To validate this, we compared float32 and float64 simulations over a representative portion of the evolution and confirmed that all physically relevant observables agree to within our target accuracy, see Sect.~\ref{app:precision}. One potential concern with float32 is the accumulation of rounding errors over many time steps; we show explicitly that this does not occur here, and that single precision remains reliable throughout the full simulation window.
 
 
After all optimizations on a single GPU have been exhausted, one can still push the numerics by using multiple GPUs. In this case, we distributed the batches of matrix multiplications across the devices using a greedy load-balancing rule that aims to maximize the utilization of each device, while performing the SVDs on the CPU using 16 threads. Furthermore, only the active local tensors, environments, effective Hamiltonian data, and current Krylov vector are staged to the GPU devices to preserve memory, while the CPU acts as the storage device. Our largest simulation at $\chi \approx 62{,}000$ ($\epsilon_{\mathrm{tol}} = 10^{-6}$, \texttt{float32}) took 16h\,12m until $t=7$. See Table~\ref{tab:runtime_to_t7} for runtimes of all our experiments.

\subsection{SVD truncation threshold}

For a global quench, the entanglement entropy is believed to grow linearly $S\lr{t}=a\cdot t$~\cite{calabrese2005evolution}, which means that the bond dimension should grow exponentially $\chi\lr{t} \sim \exp\lr{a t}$.
TDVP achieves this growth by doing 2-site updates, which increases the maximum possible number of states on the link, and then keeping the most relevant states using SVD.
 
\begin{figure}[htbp]
\includegraphics[width=\columnwidth]{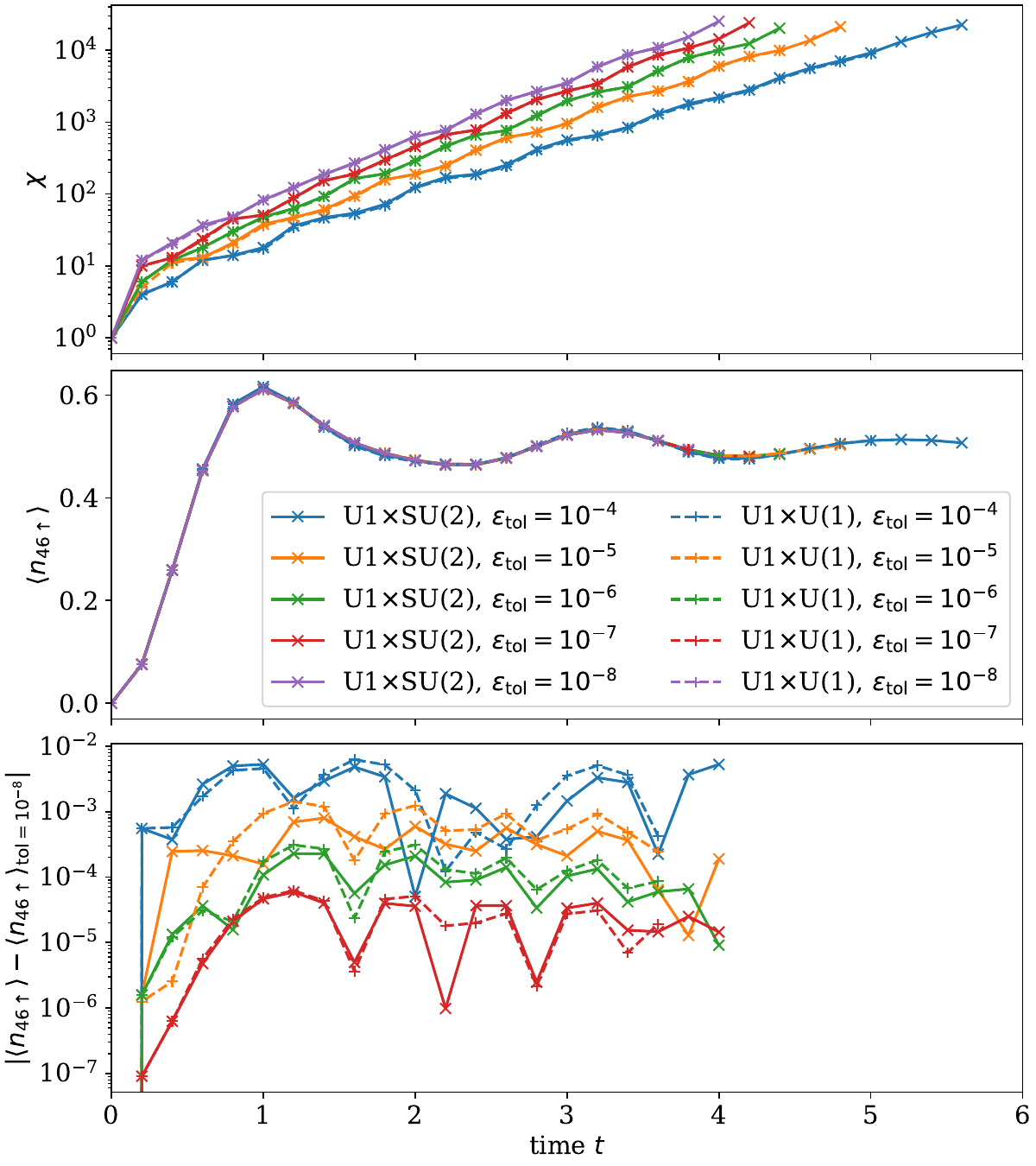}
\caption{\label{fig:tol}
SVD tolerance analysis with unrestricted bond dimension. Top: exponential
$\chi(t)$ growth for different $\epsilon_\mathrm{tol}$. Middle: the
observable $\avg{n_{46\uparrow}}$ is nearly identical across tolerances.
Bottom: deviation from the most accurate result
($\epsilon_\mathrm{tol}=10^{-8}$).
}
\end{figure}

An important hyperparameter is the compression tolerance, i.e. the threshold for the sum of discarded singular values (``truncated weight''), which regulates the growth rate. Setting it too large will not faithfully follow the state, while setting it too small will lead to unnecessarily expensive computations. In Fig.~\ref{fig:tol}, we vary the compression tolerance across several orders of magnitude without capping the bond dimension and propagating until the numerical resources are exhausted, while looking at the observable $\avg{n_{46\uparrow}}$. We find that for the given problem, even a tolerance of $10^{-4}$ is sufficient to capture the observable at least until $t\sim 5$, and this is the value we use in the plots of the main text.
 
\begin{figure}[htbp]
\includegraphics[width=\columnwidth]{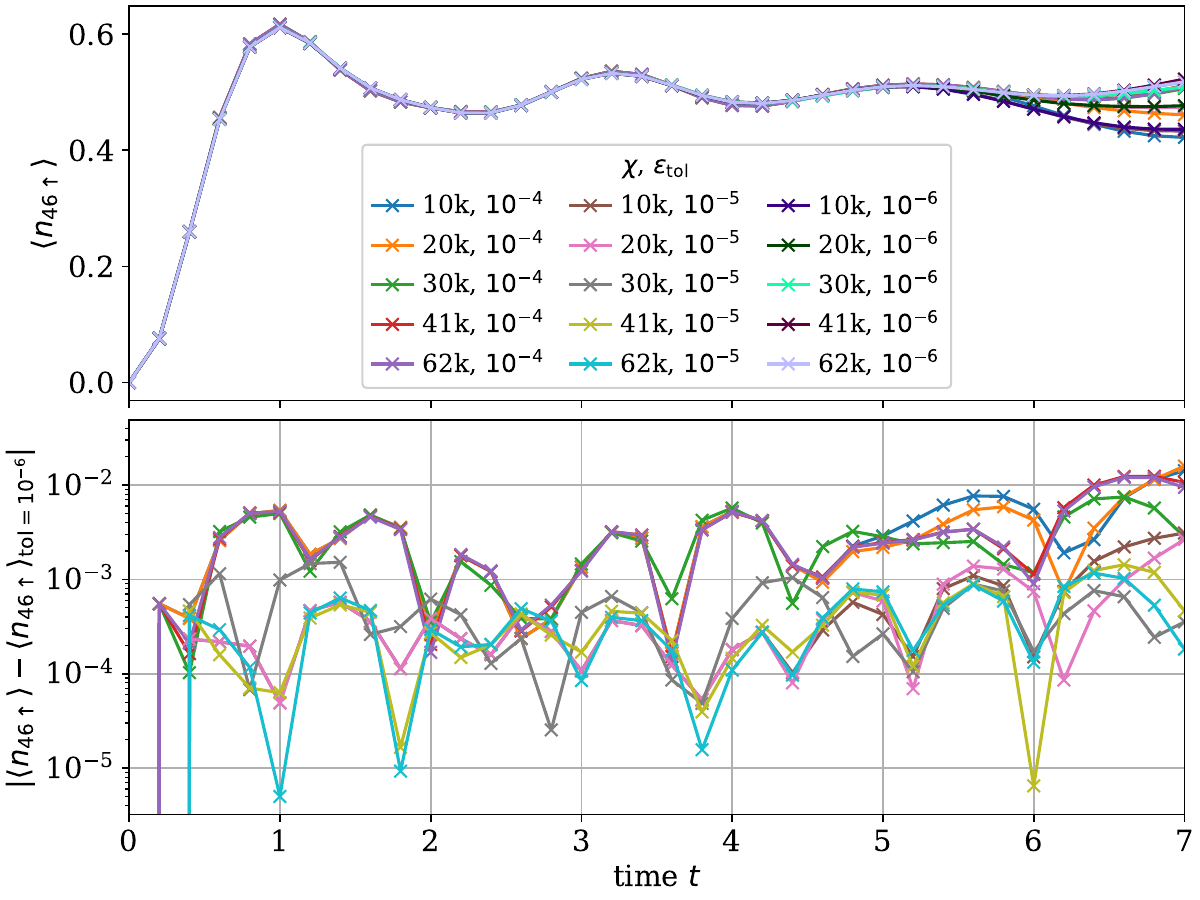}
\caption{\label{fig:tolcheck}
Tolerance comparison with capped bond dimension and the adaptive
1-site/2-site switch. Top: $\avg{n_{46\uparrow}}$ for each
($\chi$, $\epsilon_\mathrm{tol}$) combination.
Bottom: deviation from the $\epsilon_\mathrm{tol}=10^{-6}$ reference.
}
\end{figure}

Note that we would still like to propagate further with a capped bond dimension beyond this.
Even though the exponential growth of $\chi$ cannot be followed further, it does not mean that the results immediately become invalid. To get a sense of the errors in this case, we compare the results with $\epsilon_{\text{tol}}=10^{-4}, 10^{-5}$ to the most accurate result at $\epsilon_{\text{tol}}=10^{-6}$ in Fig.~\ref{fig:tolcheck} and find that the error in the observable remains at most of order $O(10^{-2})$.

\subsection{Float32 vs.\ float64 validation}\label{app:precision}

\begin{figure}[htbp]
\includegraphics[width=\linewidth]{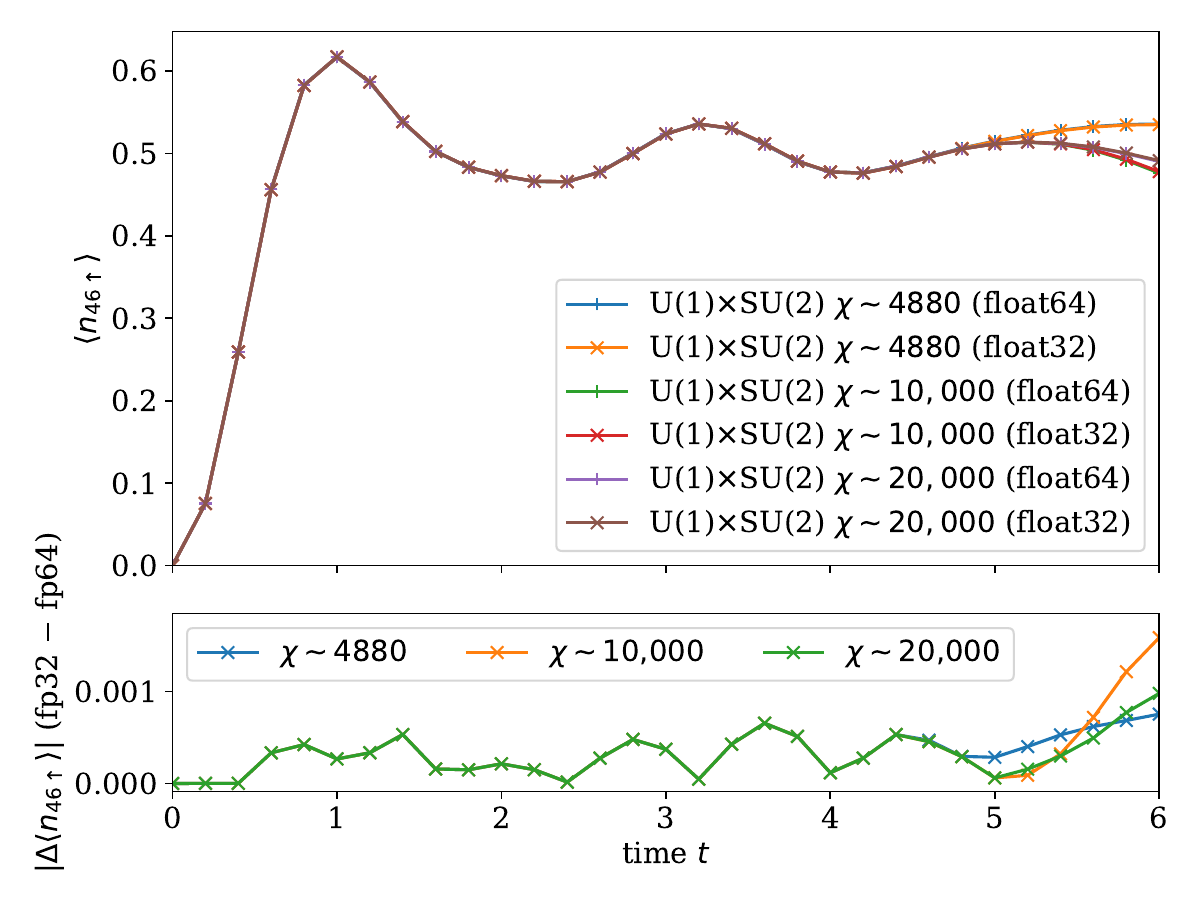}
\caption{\label{fig:float32}
Float32 vs.\ float64 comparison at $\epsilon_\mathrm{tol}=10^{-4}$.
Top: $\avg{n_{46\uparrow}}$ for both precisions.
Bottom: absolute difference, remaining $O(10^{-3})$ or smaller throughout.
}
\end{figure}

The noise floor of the Q-CTRL quantum circuits substantially exceeds the
rounding errors of \texttt{float32} arithmetic, so single precision suffices
for benchmarking purposes. Figure~\ref{fig:float32} compares
\texttt{float32} and \texttt{float64} evolutions; the absolute deviation in
$\avg{n_{46\uparrow}}$ remains below $10^{-3}$ throughout. Using
\texttt{float32} also yields a modest $6$--$23\%$ wall-clock speedup and
halves GPU memory consumption, enabling the largest bond dimensions.

\section{Density profile comparison}\label{app:missing}
 
\begin{figure*}[htbp]
\includegraphics[width=0.9\textwidth]{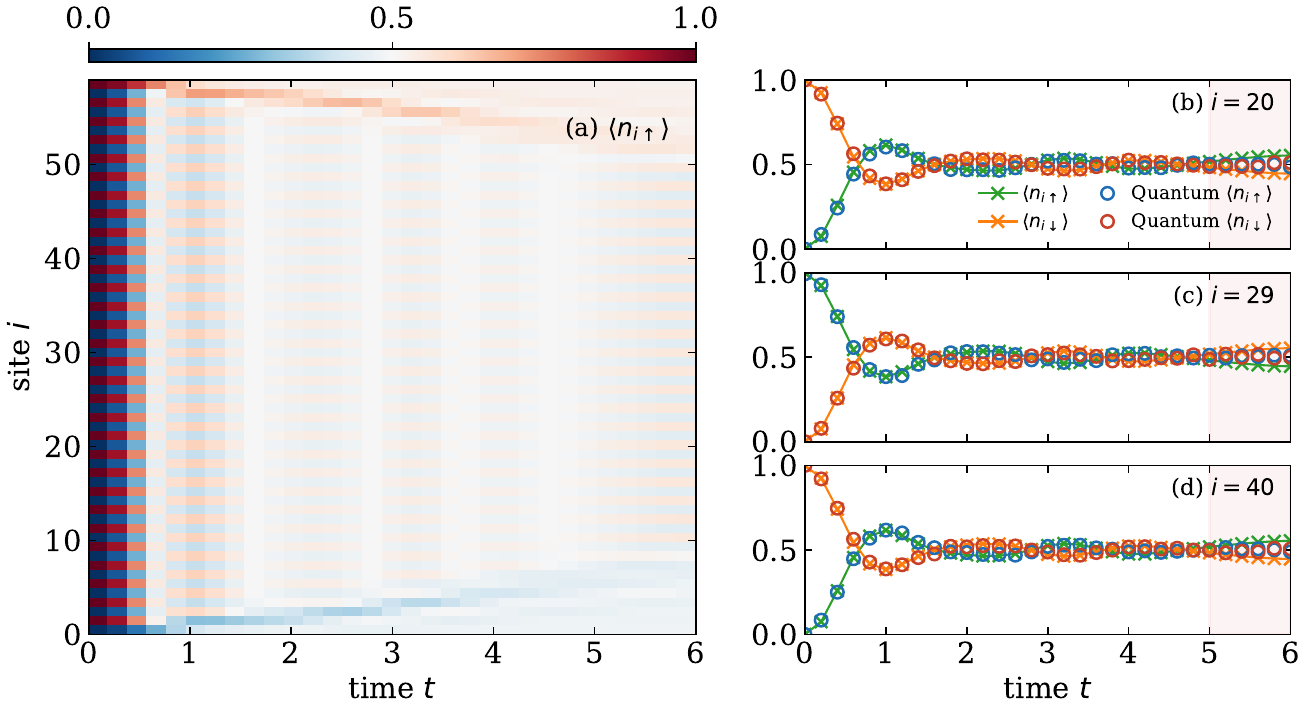}
\caption{\label{fig:profile_u1xu1_4096}
Density heatmap as in Fig.~\ref{fig:heatmap} but using
$\mathrm{U}(1)\times\mathrm{U}(1)$ symmetry at $\chi=4096$,
$\epsilon_\mathrm{tol}=10^{-4}$. Visible deviations at late times confirm
that this bond dimension is insufficient to benchmark the quantum results (area shaded red).
}
\end{figure*}

\begin{figure*}[htbp]
\includegraphics[width=0.9\textwidth]{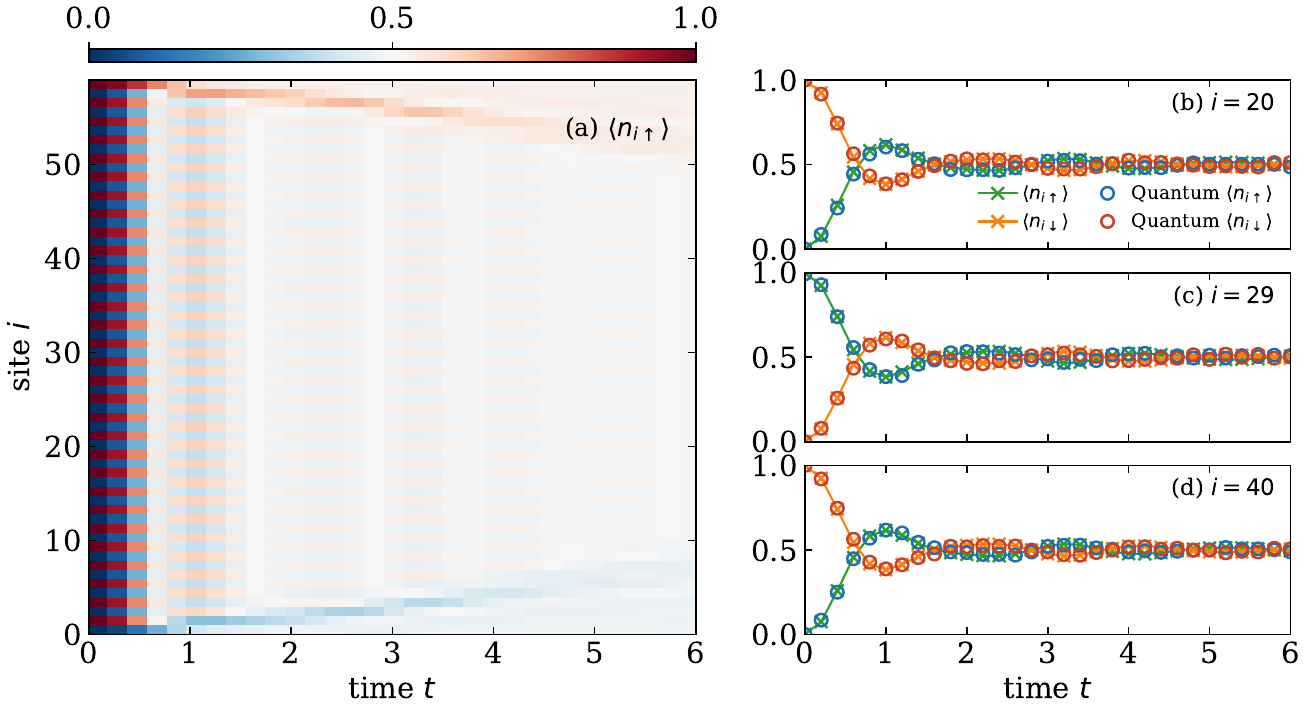}
\caption{\label{fig:res_heatmap_12288}
Density heatmap as in Fig.~\ref{fig:heatmap}, but computed with
$\mathrm{U}(1)\times\mathrm{SU}(2)$ symmetry at $\chi\approx30{,}000$ and
$\epsilon_{\mathrm{tol}}=10^{-4}$. These parameters are sufficient for convergence
up to $t=6$.}
\end{figure*}

Figure~\ref{fig:profile_u1xu1_4096} shows the full density profile computed with
$\mathrm{U}(1)\times\mathrm{U}(1)$ symmetry at $\chi=4096$ and
$\epsilon_{\mathrm{tol}}=10^{-4}$ --- a comparison absent from the Q-CTRL
paper~\cite{qctrl2025fermihubbard}. The profiles deviate visibly at late times,
confirming that this bond dimension is insufficient for convergence. Switching to
$\mathrm{U}(1)\times\mathrm{SU}(2)$ symmetry, convergence is achieved at
$\chi\approx30{,}000$ (Fig.~\ref{fig:res_heatmap_12288}).

\section{Supplementary Figures and Tables}

\begin{table*}[htbp]
  \centering
  \begin{tabular*}{\textwidth}{@{\extracolsep{\fill}} l l l r r r r r}
    \toprule
    $\epsilon_{\text{tol}}$ & Symmetry & Prec. & $\chi_{\mathrm{SU}(2)}$ & $\chi$ & $N_{\mathrm{GPU}}$ & $t=6$ & $t=7$ \\
    \midrule
    $10^{-4}$ & U(1)$\times$U(1) & fp64 & --- & 4\,096 & 1 & 3h\,27m & --- \\
     & U(1)$\times$SU(2) & fp64 & 2\,048 & 4\,880 & 1 & 1h\,51m & --- \\
     & U(1)$\times$SU(2) & fp32 & 2\,048 & 4\,880 & 1 & 1h\,40m & --- \\
     & U(1)$\times$SU(2) & fp64 & 4\,096 & 10\,000 & 1 & 2h\,30m & 2h\,42m \\
     & U(1)$\times$SU(2) & fp64 & 8\,192 & 20\,000 & 1 & 3h\,33m & 3h\,51m \\
     & U(1)$\times$SU(2) & fp32 & 12\,288 & 30\,000 & 1 & 4h\,06m & 4h\,27m \\
     & U(1)$\times$SU(2) & fp32 & 16\,384 & 41\,000 & 2 & 5h\,31m & 8h\,20m \\
     & U(1)$\times$SU(2) & fp32 & 24\,576 & 62\,000 & 4 & 5h\,21m & 13h\,57m \\
    \midrule
    $10^{-5}$ & U(1)$\times$U(1) & fp64 & --- & 4\,096 & 1 & 3h\,51m & --- \\
     & U(1)$\times$SU(2) & fp64 & 2\,048 & 4\,880 & 1 & 2h\,15m & --- \\
     & U(1)$\times$SU(2) & fp32 & 2\,048 & 4\,880 & 1 & 2h\,49m & --- \\
     & U(1)$\times$SU(2) & fp64 & 4\,096 & 10\,000 & 1 & 2h\,53m & 3h\,06m \\
     & U(1)$\times$SU(2) & fp64 & 8\,192 & 20\,000 & 1 & 3h\,29m & 3h\,49m \\
     & U(1)$\times$SU(2) & fp32 & 12\,288 & 30\,000 & 1 & 5h\,26m & 5h\,48m \\
     & U(1)$\times$SU(2) & fp32 & 16\,384 & 41\,000 & 2 & 7h\,13m & 8h\,11m \\
     & U(1)$\times$SU(2) & fp32 & 24\,576 & 62\,000 & 4 & 13h\,33m & 15h\,40m \\
    \midrule
    $10^{-6}$ & U(1)$\times$U(1) & fp64 & --- & 4\,096 & 1 & 3h\,59m & --- \\
     & U(1)$\times$SU(2) & fp64 & 2\,048 & 4\,880 & 1 & 2h\,02m & --- \\
     & U(1)$\times$SU(2) & fp64 & 4\,096 & 10\,000 & 1 & 2h\,49m & 3h\,02m \\
     & U(1)$\times$SU(2) & fp64 & 8\,192 & 20\,000 & 1 & 3h\,32m & 3h\,53m \\
     & U(1)$\times$SU(2) & fp32 & 12\,288 & 30\,000 & 1 & 4h\,28m & 4h\,51m \\
     & U(1)$\times$SU(2) & fp32 & 16\,384 & 41\,000 & 2 & 7h\,24m & 8h\,21m \\
     & U(1)$\times$SU(2) & fp32 & 24\,576 & 62\,000 & 4 & 14h\,08m & 16h\,12m \\
    \bottomrule
  \end{tabular*}
  \caption{Wall-clock runtime to reach $t=6$ and $t=7$ as a function of
  convergence tolerance $\epsilon_\mathrm{tol}$ and bond dimension.
  Runtimes are non-monotonic because the algorithm adaptively switches
  from 2-site to 1-site TDVP once the bond dimension saturates.}
  \label{tab:runtime_to_t7}
\end{table*}

\begin{table}[htbp]
\centering
\caption{Irrep degeneracies $d_{n,j}$ for the 32 symmetry sectors present
at $t=6$ in our $\mathrm{U}(1)\times\mathrm{SU}(2)$ TDVP evolution at
$\chi_{\mathrm{SU(2)}}=12{,}288$ ($\chi\approx30{,}000$),
for the left bond of the MPS tensor at site 30.}
\label{tab:irrep_degeneracies}
\begin{tabular}{ccc@{\hskip 1em}|@{\hskip 1em}ccc}
\toprule
$n$ & $j$ & $d_{n,j}$ & $n$ & $j$ & $d_{n,j}$ \\
\midrule
$-6$ & $0$           &    8 & $ 0$ & $0$           & 1097 \\
$-6$ & $1$           &    1 & $ 0$ & $1$           & 1387 \\
$-5$ & $\frac{1}{2}$ &   86 & $ 0$ & $2$           &  287 \\
$-5$ & $\frac{3}{2}$ &   13 & $ 0$ & $3$           &    8 \\
$-4$ & $0$           &  241 & $ 1$ & $\frac{1}{2}$ & 1157 \\
$-4$ & $1$           &  233 & $ 1$ & $\frac{3}{2}$ &  494 \\
$-4$ & $2$           &   19 & $ 1$ & $\frac{5}{2}$ &   42 \\
$-3$ & $\frac{1}{2}$ &  799 & $ 2$ & $0$           &  382 \\
$-3$ & $\frac{3}{2}$ &  307 & $ 2$ & $1$           &  421 \\
$-3$ & $\frac{5}{2}$ &   10 & $ 2$ & $2$           &   67 \\
$-2$ & $0$           &  920 & $ 3$ & $\frac{1}{2}$ &  178 \\
$-2$ & $1$           & 1135 & $ 3$ & $\frac{3}{2}$ &   58 \\
$-2$ & $2$           &  213 & $ 4$ & $0$           &   35 \\
$-1$ & $\frac{1}{2}$ & 1735 & $ 4$ & $1$           &   29 \\
$-1$ & $\frac{3}{2}$ &  837 & $ 5$ & $\frac{1}{2}$ &    8 \\
$-1$ & $\frac{5}{2}$ &   80 & $ 6$ & $0$           &    1 \\
\bottomrule
\end{tabular}
\end{table}

\end{document}